# Constant Time Encryption as a Countermeasure against Remote Cache Timing Attacks

Darshana Jayasinghe, Roshan Ragel and Dhammika Elkaduwe
Department of Computer Engineering
University of Peradeniya
Peradeniya 20400 Sri Lanka

*Abstract*- **Rijndael was standardized in 2001 by National Institute of Standard and Technology as the Advanced Encryption Standard (AES). AES is still being used to encrypt financial, military and even government confidential data. In 2005, Bernstein illustrated a remote cache timing attack on AES using the client-server architecture and therefore proved a side channel in its software implementation. Over the years, a number of countermeasures have been proposed against cache timing attacks both using hardware and software. Although the software based countermeasures are flexible and easy to deploy, most of such countermeasures are vulnerable to statistical analysis. In this paper, we propose a novel software based countermeasure against cache timing attacks, known as constant time encryption, which we believe is secure against statistical analysis. The countermeasure we proposed performs rescheduling of instructions such that the encryption rounds will consume constant time independent of the cache hits and misses. Through experiments, we prove that our countermeasure is secure against Bernstein's cache timing attack.**

## I. Introduction

The Caesar cipher is named after Julius Caesar, in which each letter in the plaintext is replaced by another letter some fixed number of positions down the alphabet to protect messages of military significance [7]. Starting from Caesar cipher cryptography has come a long way with the invention of Computers and the Internet. Today cryptography holds an important role since a large amount of military and financial data are being transferred through the Internet, an unsecure media, as digital data.

In 1976 National Bureau of Standards (NBS) selected the Data Encryption Standard (DES), which is a block cipher, as an official Federal Information Processing Standard (FIPS) for United States. DES used 56 bits key size that became insecure with the increasing processing power of the computers. In 1999 *distributed.net* and Electronic Frontier Foundation in collaboration publicly broke a DES key within 22 hours and 15 minutes. In addition, in 1996, Paul C. Kocher published a paper [1], which illustrated a timing attack for public key algorithms.

In 1996, National Institute of Standards and Technology (NIST), previously known as NBS, called for new encryption algorithms to be the next generation encryption algorithm. In November 2001 after a 5 year of standardization, Rijndael was chosen to be the next generation encryption standard [2], formally named AES (Advanced encryption standard). Today, most of the military, financial, and government confidential data are being encrypted by AES [3] and therefore it plays a major role in cyber security.

Unlike DES, AES does not use a Feistel structure; instead, it uses a substitution and permutation network that is fast in hardware as well as in software. AES can encrypt a 128-bit block at a time, which is represented as a 16-byte array (4x4) referred as a state array. As depicted in Figure 1, AES is composed of the following four steps: *sub byte, shift rows, mix columns* and *add round key*, which are used to build the cipher from a plain text.

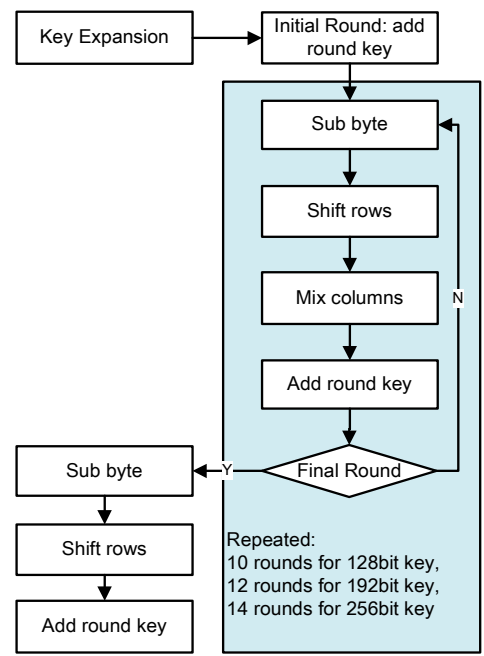

Figure 1. The AES Algorithm as a Block Diagram

AES supports three key sizes 128 bit, 192 bit and 256 bit and differing number of rounds based on the key size 10, 12 and 14 respectively. In the software implementation of AES, to speed up the encryption process, sub byte, shift columns and mix rows calculations are pre-calculated and are made as table (T tables) lookups. There are four T tables with each the size of 1 KB. Typically the use of T tables in the software implementation of AES will increase the encryption performance by 60 times compared to the one without using them.

An attacker uses cryptanalysis to study the methods of obtaining the valuable information hidden in the cipher text. It is a vast study, which varies from frequency analysis to rubber horse analysis. One prominent cryptanalysis is the side channel attack [4]. The side channel attack is a crypto attack that is performed on the physical implementation of a crypto system. For example timing information [5], power consumption, electromagnetic radiation and even the cache content are considered side channels.

When NIST analysed AES for security vulnerabilities, they considered data dependent rotations and branching conditions as the only side channels. However using the timing variations caused by cache hits and misses due to T tables, Daniel Bernstein illustrated a remote cache timing attack on AES in April 2005 [6]. In October 2005, Osvik et al. presented a paper demonstrating several cache timing attacks against AES, where one of their attacks could break the AES in 65 milliseconds with 800 encryptions [7].

In 2009 July Biryukov and Dmitry [8] described a relative key attack, which is performed on 192 bit AES and 256 bit AES. Vincent Rijmen [9] published an ironic paper on "chosen-key-relations-on the middle" attacks on AES in 2010 July.

In this paper, we consider the remote cache timing attack described by Daniel Bernstein. We demonstrate how a constant time encryption technique could be used in a software implementation of AES so that the system is secure against cache timing attacks. We demonstrate our solution on an environment as proposed by Bernstein and show that with the countermeasure proposed, AES is safe against Bernstein's cache timing attack.

The rest of the paper is organized as follows: In Section II we describe the typical countermeasures proposed against cache timing attack and group them into a number of categorises. Section III presents our methodology where we introduce the notion of constant time encryption in this papers context and how it can be applied to AES. We present the implementation of our countermeasure in Section IV where we illustrate how our methodology, constant time encryption, is applied in a software implementation of AES. In Section V, we present our results, that is how our methodology mitigates Bernstein's cache timing attack and the relevant overheads and in Section VI we conclude our paper.

## II. EXISTING COUNTERMEASURES

Since side channel attacks rely on by-products of the physical implementation of crypto systems, adding noise or covering the side channel s will eliminate the vulnerability and therefore the attack. Countermeasures prevent deducting the secret key in two ways: (1) reducing the time variance so that the attacker will not see the differences of cache hits and misses, (2) increasing the time difference so that it results in a different key space, which does not contain the secret key.

Researchers have proposed many countermeasures to prevent remote cache timing attack including, loading T tables into registers [4], adding hardware components to calculate T tables [4], introducing smaller T tables that would fit into the cache [2], and masking evicted timing delays from the cache. All the countermeasures have performance impact and they are investigated in [6]. Following are the descriptions of the already proposed countermeasures:

A. *Loading T tables in to the registers:* If the CPU register file is large such that it can hold four T tables, the T tables can be accessed with a constant time. However, commonly used hardware architectures (such as x86 or ARM) do not have enough registers to hold all four T tables. Even worse, making your own "special" register file in hardware is expensive as they will consume a large portion of your processor floor space.

B. *Adding hardware components to calculate T tables*: If the functional units necessary for computing T table values are embedded into the core or die as a hardware component, then the T table lookups are not necessary. Intel has implemented this feature in new processor family (i.e. core i5 and core i7). However, such components are going to consume lot of logic gates and therefore not feasible in the current embedded microprocessors.

C. *Smaller T tables that would fit into the cache*: Either one has to start from S boxes or some intermediate table the size of which is in between S box Tables and T tables and calculate the rest. T tables were introduced to speed up the software encryption process. Therefore, using a smaller table and calculating rest will have a huge performance penalty. In addition, if the intermediate table does not fit in to the cache with the encryption data, then the vulnerability will still be there.

D. *Masking evicted timing information from the cache:* Adding random loops, random sleeps in the middle of the encryption will cause false timing details that will lead to a wrong key space. From the previous experiments, it has been proved that this has the least performance impact. However, such masking can be vulnerable to the statistical analysis; that is, by repeating the attack many times one can remove the random noise using statistical methods. Therefore, the vulnerability can be still there.

E. *Pre-fetching:* Pre-fetching can be applicable to either hardware or operating system. If the relevant T tables can be fetched before it actually is needed, it will cause no cache misses. To perform hardware pre-fetching, special hardware is needed which are not provided in microprocessors that come with day-to-day computers.

F. *Disabling the cache:* If the cache in the system is disabled, then the RAM memory of the system has to be utilised to store the T tables. The RAM (Random Access Memory) is large enough to hold all the T tables. Therefore, every access will take the same amount of time, thus removing the vulnerability (we ignore operating system's memory paging for now). However, since the RAM is far slower than the cache there will be a huge performance penalty. Moreover, disabling the cache should be supported by the hardware and the operating system kernel. Only a few desktop motherboards support disabling the L2 cache and typical operating systems do not allow cache disabling.

### III. CONSTANT TIME ENCRYPTION

We define a cryptosystem to have constant time encryption, when it consumes equal amount of time for all encryption rounds independent of the data being encrypted. Typically, achieving constant time encryption could take two paths: one is to deploy a cache that is large enough to hold all four T tables (plus the rest of the data) and therefore producing all cache hits and the other is to have a system without a cache where the performance of the system will be degraded due to high latency memory accesses. In both cases, the time taken to access data from different parts of the memory or cache will be a constant and therefore constant time encryption.

In this paper, we take a third approach, where we maintain the same cache structure of the system (that is, we propose not to do any hardware changes). However, we achieve constant time encryption via scheduling the instructions such that the encryption rounds will circumvent delays due to cache misses. Let us look at an example. Assume a cache hit latency of 2 clock cycles, a miss latency of 6 clock cycles and instruction execution latency of 1 clock cycle. Figure 2 depicts an instruction sequence (Figure 2a) and how it can be rescheduled to avoid time delay due to cache misses (Figure 2b).

In Figure 2a, the second instruction in the sequence is a load instruction and instruction number 6 uses the data loaded in instruction 2. When the load instruction encounters a cache hit, the data loaded will be available from instruction 3. However, when there is a cache miss for the load instruction, the data loaded will only be available from instruction 8.

Therefore, although a cache hit will not delay the execution of instruction 6, a cache miss will. This is the fact that causes the time difference and therefore the vulnerability. As shown in Figure 2b, if instruction 6 can be rescheduled to the eighth location (let us assume that instructions e and f do not depend on instruction 6), then for both a cache miss and a cache hit the use data instruction is going to behave the same way achieving constant time encryption.

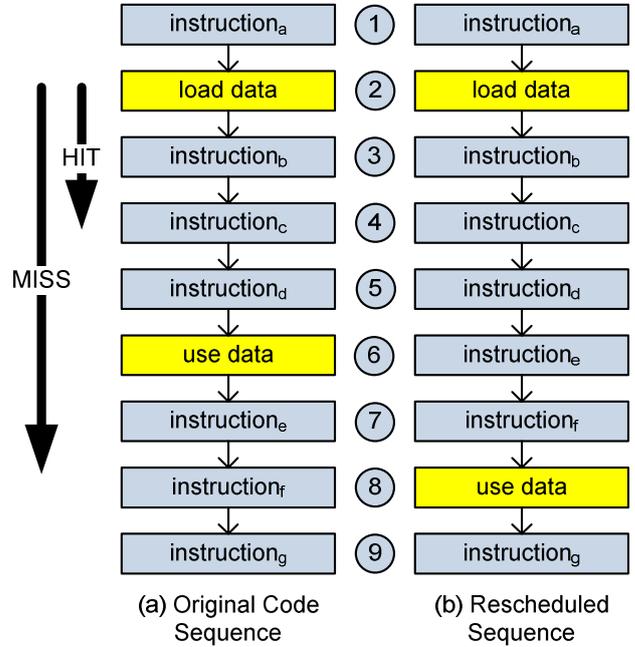

Figure 2. Rescheduling for Constant Time Encryption

Details of how the idea presented in this section on constant time encryption is applied to the software implementation of AES are discussed in the following section.

### IV. SCHEDULING THE AES INSTRUCTIONS

OpenSSL code for AES encryption is written in C and it is organized as shown in Figure 3.

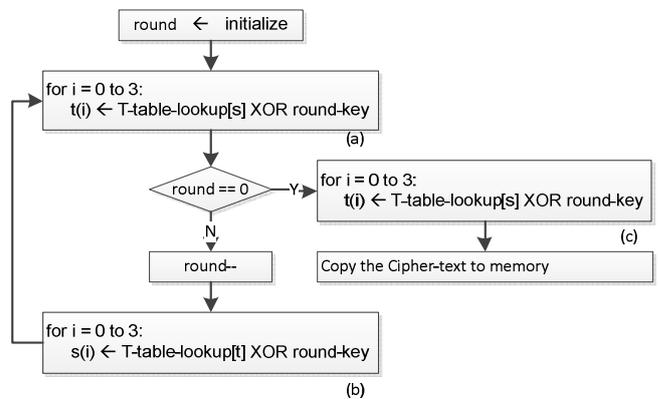

Figure 3. Block Diagram of OpenSSL AES Code

Pre-calculated variables of s are used at the start in block (a) then the updated values of *t(s)* are used to calculate variables of *s* again. At last, the final round is executed and the cipher text is copied into memory. The rest of this section illustrates how the manual scheduling of the AES OpenSSL instructions can be done with examples. Our manual intervention to the compiler supported scheduling has three major steps and they are discussed in the rest of this section.

*A. Decomposing the OpenSSL AES encryption code into smaller bitwise operations.*

For example the code segment shown in Figure 4, which is part of the OpenSSL AES encryption code, can be decomposed into smaller bitwise operations as shown in Figure 5. Such decomposition is typically part of the instruction scheduling process of a compiler.

$$t0 = \\ Te0[(s0 >> 24)\,] \wedge \\ Te1[(s1 >> 16)\,\&0xff] \wedge \\ Te3[(s3)\,\&0xff\,] \wedge \\ rk[4];$$

Figure 4. A Sample Source Code from OpenSSL AES Encryption

During our manual scheduling process, we performed this decomposition ourselves so that we will be able to apply the rest of the steps without modifying the compiler. However, such changes can be integrated into a compiler, so that the compiler will then be able to perform "side-channel aware" scheduling.

$$u00 = s0; \\ u00 = u00 >> 24; \\ v00 = Te0[u00]; \\ t0 = rk[4] \wedge v00;$$

$$u01 = s1; \\ u01 = u01 >> 16; \\ u01 = u01\,\&\,0xff; \\ v01 = Te1[u01]; \\ t0 = t0 \wedge v01;$$

$$u02 = s2; \\ u02 = u02 >> 8; \\ u02 = u02\,\&\,0xff; \\ v02 = Te2[u02]; \\ t0 = t0 \wedge v02;$$

$$u03 = s3; \\ u03 = u02\,\&\,0xff; \\ v03 = Te3[u3]; \\ t0 = t0 \wedge v03;$$

Figure 5. Decomposed Bitwise Operations for Code Segment in Figure 4

In Bernstein's attack, the AES source codes are compiled in gcc with "–O3" optimization. However due to the manual scheduling we would like to perform and to make sure that the compiler does not undo our scheduling, we will not use gcc's optimizations. As mentioned earlier, the decomposed program for the code segment in Figure 4 will look like the one in Figure 5. Intermediate variables are used to hold the data temporally. The decomposition we perform and not using the complier's optimization options would increase the code size by little and will also increase the execution penalty.

$$u01 = s1; \\ u02 = s2; \\ u00 = s0; \\ u01 = u01 >> 16; \\ u02 = u02 >> 8; \\ u03 = s3; \\ u10 = s1; \\ u11 = u11 >> 16;$$

$$u00 = u00 >> 24; \\ u01 = u01\,\&\,0xff; \\ u02 = u02\,\&\,0xff; \\ u03 = u03\,\&\,0xff; \\ u10 = u10 >> 24; \\ u11 = u11\,\&\,0xff;$$

$$v00 = Te0[u00]; \\ u01 = u01\,\&\,u01; \\ u02 = u02\,\&\,u02; \\ u03 = u03\,\&\,u03; \\ v10 = Te0[u10]; \\ u11 = u11\,\&\,u11;$$

$$t0 = rk[4] \wedge v00; \\ v01 = Te1[u01]; \\ u02 = u02\,\&\,u02; \\ u03 = u03\,\&\,u03; \\ t1 = rk[5] \wedge v10; \\ v11 = Te1[u11];$$

Figure 6. The Code Segment Scheduled to avoid Side Channel Attack

*B. Add each and every bitwise instruction sets to queues.*

Adding the decomposed instructions to queues will make our scheduling easier to implement using a computer program and therefore the scheduling can be automated.

*C. Processing each queue.*

Each queue can be executed separately since they are data and control independent from each other. However, we want to hide the data loading time in arithmetic operations so that the instructions can be scheduled as shown below:

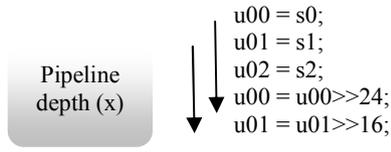

```
u00 = s0;
u01 = s1;
u02 = s2;
u00 = u00>>24;
u01 = u01>>16;
.....
```

If a processor has x number of arithmetic pipeline stages then, the second instruction of a queue can be scheduled after x number of arithmetic instructions.

When there are no arithmetic instructions to be filled in between scheduled instructions we used *asm("nop")* instruction to stall the processor. This will add a performance penalty to the encryption but will protect OpenSSL AES encryption by providing a constant time encryption. By repeating until you finish all the code segments in Figure 5, you will end up with a code segment as shown in Figure 6. This is a portion of the code we tested for a pipeline depth of 6. As shown in Figure 4, the first T table entry *Te0* is referred and it is used exactly after pipeline stage 6.

The whole process described above can be automate using a computer program and tested for performance. Since a rule based scheduling has been used to rearrange the AES encryption, the code generated by tool requires more CPU time than AES encryption code scheduled by manually.

## V. RESULTS

The efficiency and the security of the countermeasure we proposed are tested with a Pentium III CPU, which is running FreeBSD 4.8, OpenSSL 0.9.7a and gcc-2.95.3. This is the environment where Bernstein's attack was proposed and we use this environment to verify our proposed countermeasure. When compiling the AES source code, "-*O3*" optimization is not used as mentioned earlier as it will remove the *asm("nop")* instructions and might rearrange the other code segments we scheduled manually.

TABLE I. PERFORMANCE COMPARISON

| Pipeline Depth | Average Number of Cycles for encryption |
|---|---|
| 6 | 7261 |
| 8 | 7334 |
| 10 | 7384 |
| 12 | 7478 |
| 14 | 7520 |

Table I shows the clock cycle consumption of the system with the countermeasure with various pipeline depth considered. In our environment, the unprotected OpenSSL AES code takes about 6000 clock cycles. From the Bernstein's C source codes provided with paper [5], the measured time taken to encrypt a packet is not just the time to encrypt the packet. It is the time taken to execute from line 2 to line 8 as shown in Figure 7.

```
1: for(i = 0; i < 40; + + i)out[i] = 0;
2: *(unsigned int*)(out + 32) = timestamp();
3: if(len < 16)return;
4: for(i = 0; i < 16; + + i)out[i] = in[i];
5: for(i = 16; i < len; + + i)workarea[i] = in[i];
6: AES_encrypt(in, workarea, &expanded);
7: for(i = 0; i < 16; + + i)out[16 + i] = scrambledzero[i];
8: *(unsigned int*)(out + 36) = timestamp();
```

Figure 7. Measuring Clock Cycles

Therefore, the clock cycle values mentioned in Table I are not just the time taken only for AES encryption. It takes only about 500-600 numbers of cycles to execute the AES encryption code on the experimental computer we configured and therefore we have scheduled a small portion compared to the total number of cycles (approximately 10%).

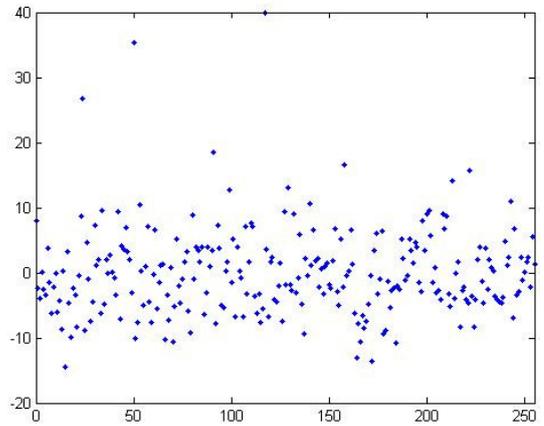

Figure 8. Timing Data for Unprotected OpenSSL

As shown in Figure 8, the unprotected OpenSSL clearly has timing data evicted during the encryption process. The graph in Figure 8 is produced for $11^{th}$ key byte. The horizontal axis is the choice of n[11], from 0 through 255. The vertical axis is the average number of clock cycles for a large number of 800-byte packets with that choice of n[11], minus the average number of cycles for all choices of n[11]. It is clearly seen that some possibilities such as n[11]=52 and n[11]=30 make the average which was calculated, to deviate from the average. This information leakage will be a good side channel for the attacker and therefore the software implementation of AES becoming vulnerable.

Figure 9 is drawn for the timing data for the protected OpenSSL AES encryption process. All the data lie in a small region of +1.5 to -3.0. Therefore the timing data is not visible enough for the attacker to deduce the secret key. However, in the previous scenario timing details are spread across the graph. One can conclude reducing the margin will help to find the secret key. But reducing the margin of the correlation program will increase the number of possibilities of the key space. Therefore the attacker will not get much gain by reducing the margin. Since we were unable to find the suitable pipe line stages of a Pentium III processor, the efficacy of the countermeasure is tested for various number of pipeline stages.

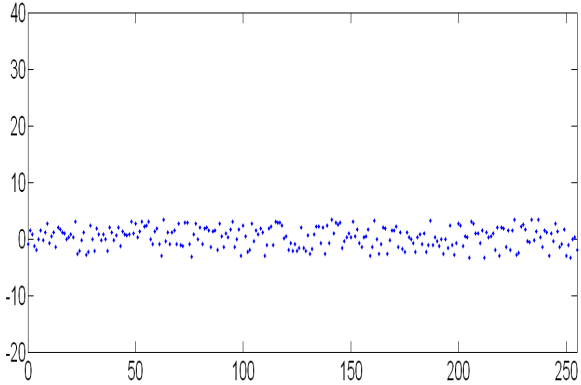

Figure 9. Timing Data for Protected OpenSSL with the Countermeasures

Table III shows the performance overhead due to the countermeasure proposed against various number of pipeline depths considered. According to Table II the variation of the performance degradation is very small as the pipeline depth changes. Therefore, we can conclude that all the pipeline depths have almost same efficiency, because the difference between 1.21 and 1.25 is small when it is considered in the practical scenario.

TABLE II. TOTAL OVERHEAD

| Pipeline Depth | Performance degradation compared to unprotected AES (Times) |
|---|---|
| 6 | 1.21 |
| 8 | 1.22 |
| 10 | 1.23 |
| 12 | 1.23 |
| 14 | 1.25 |

Through Bernstein's attack, we checked the efficacy of our countermeasure. That is, we put the AES implementation with the countermeasure incorporated under test. Table III shows the key space what was discovered by Bernstein's attack on the OpenSSL AES implementation that was protected with our countermeasure. From Table III when the pipeline depth is 12 it is reported that the maximum key space of $3.4 \times 10^{38}$ which is almost equal to the brute force attack. When then pipeline depth is 10 it gives a small key space compared to other key spaces but it is more likely a deviated key space from the correct one. Therefore, we would conclude that out of the considered pipeline depths, 12 have given the best countermeasure.

TABLE III. TOTAL KEY SPACE WITH THE COUNTERMEASURE

| Pipeline Depth | Key Space |
|---|---|
| 6 | 3.22 E 37 |
| 8 | 3.22 E 37 |
| 10 | 2.00 E 33 |
| 12 | 3.40 E 38 |
| 14 | 1.33 E 18 |

VI. CONCLUSIONS

Previous researches proved that the famous block cyber, AES, is vulnerable to cache timing attack, a famous side-channel attack. Typical software based countermeasures proposed in the past against cache timing attack are vulnerable to statistical analysis. In this paper, we propose a software based countermeasure known as constant time encryption that we believe is secure against cache timing attack including statistical analysis. The constant time encryption proposed makes sure that the encryption rounds take same amount of clock cycles independent of the cache hits and misses, eliminating the side-channel vulnerability existed in the past. However, our countermeasure is going to burden the programmer with additional challenges such as identifying the pipeline depth of a processor and programming accordingly and the changes we make to the software code will make it non-portable.